\documentclass[12pt]{article}
\usepackage{epsfig}
\usepackage{aaspp4}

\def\spose#1{\hbox to 0pt{#1\hss}}
\newcommand\lsim{\mathrel{\spose{\lower 3pt\hbox{$\mathchar"218$}}
     \raise 2.0pt\hbox{$\mathchar"13C$}}}
\newcommand\gsim{\mathrel{\spose{\lower 3pt\hbox{$\mathchar"218$}}
     \raise 2.0pt\hbox{$\mathchar"13E$}}}
\def\arcmin{\hbox{$^\prime$}}

\def\ale{\mathrel{\hbox{\rlap{\hbox{\lower4pt\hbox{$\sim$}}}\hbox{$<$}}}}
\def\age{\mathrel{\hbox{\rlap{\hbox{\lower4pt\hbox{$\sim$}}}\hbox{$>$}}}}

\begin{document}

\title{Optical and Radio Observations of the Afterglow from GRB~990510: Evidence for a Jet}

\author{F. A. Harrison\altaffilmark{1}, J. S. Bloom\altaffilmark{1}, 
D.  A. Frail\altaffilmark{2}, R. Sari\altaffilmark{3},
S. R. Kulkarni\altaffilmark{1}, S. G. Djorgovski\altaffilmark{1},
T. Axelrod\altaffilmark{4}, J. Mould\altaffilmark{4},
B. P. Schmidt\altaffilmark{4}, M. H. Wieringa\altaffilmark{6}, R. M.
Wark\altaffilmark{6}, R. Subrahmanyan\altaffilmark{6},
D. McConnell\altaffilmark{6}, P. J. McCarthy\altaffilmark{5},
B. E. Schaefer\altaffilmark{7}, R. G. McMahon\altaffilmark{8},
R. O. Markze\altaffilmark{5}, E. Firth\altaffilmark{8},
P. Soffitta\altaffilmark{9}, and L. Amati\altaffilmark{10}}

\altaffiltext{1}{Palomar Observatory 105-24, California Institute of Technology, 
Pasadena, CA 91125} 
\altaffiltext{2}{National Radio Astronomy Observatory, Socorro, NM 87801}
\altaffiltext{3}{California Institute of Technology, Theoretical
Astrophysics 103-33, Pasadena, CA 91125} 
\altaffiltext{4}{Research School of Astronomy, Australian National University,
Private Bag, Weston Creek P.O., ACT 2611, Australia}
\altaffiltext{5}{Observatories of the Carnegie Institute of Washington, 
813 Santa Barbara Street, Pasadena, CA 91101-1292}
\altaffiltext{6}{Australian Telescope National Facility, CSIRO, Epping, NSW 2121, Australia}
\altaffiltext{7}{Department of Physics, Yale University, New Haven, CT 06520}
\altaffiltext{8}{Institute of Astronomy, Madingley Road, Cambridge CB3 OHA UK}
\altaffiltext{9}{Istituto Astrofisica Spaziale, CNR, 
Area di Ricerca Tor Vergata, Via Fosso del Cavaliere 100, 00133 Roma, Italy}
\altaffiltext{10}{Istituto Tecnologie e Studio Radiazioni Extraterrestri, 
CNR, Via Gobetti 101, 40129 Bologna, Italy}

\begin{abstract}
We present multi-color optical and two-frequency radio observations of
the bright SAX event, GRB~990510.  Neither the well-sampled optical
decay, nor the radio observations are consistent with simple spherical
afterglow models.  The achromatic steepening in the optical band and
the early decay of the radio afterglow, both occuring at $t \sim
1$~day, are evidence for hydrodynamical evolution of the source, and
can be most easily interpreted by models where the GRB ejecta are
collimated in a jet.  Employing a simple jet model to explain the
observations, we derive a jet opening angle of $\theta_o = 0.08
(n/1$cm$^{-3})^{1/8}$, reducing the isotropic gamma-ray energy release
of $2.9 \times 10^{53}$~erg by a factor $\sim 300$.
\end{abstract}


\section{Introduction}

Gamma-ray burst afterglow observations from X-ray through radio can be
interpreted in the context of fireball models, where a shock produced
by the interaction of relativistic ejecta with the circumburst
environment expands into the surrounding medium, producing broadband
synchrotron emission (e.g. \cite{mr97a,spn98,wax97a}).  The optical
lightcurve of GRB~970508, for example, exhibits a monotonic decay;
$F_{\nu} \propto t^{-\alpha}$ with $\alpha = 1.3$ for $\sim 200$ days
(\cite{fpg+99}), well-described by the expansion of a spherical blast
wave (\cite{wrm97}).  Recently, the rapid decay of some events has
been interpreted as evidence for jet-like, or collimated ejecta
(\cite{sph99}), but this explanation is not unique (\cite{cl99}).  For
GRB~990123, the steepening of the optical lightcurve
(\cite{kdo+99,ftm+99}) combined with the early radio decay
(\cite{kfs+99}) together provide the best evidence to-date for
deviations from spherical symmetry. Due to sparse sampling, however,
simultaneous steepening in all optical bands -- the distinctive
feature of hydrodynamic evolution of a jet -- was not clearly observed.

The bright {\em BeppoSAX} event, GRB~990510, is distinguished by
excellent sampling of the optical decay in multiple bands, and by the
early-time detection and continued monitoring of the radio afterglow.
In this Letter we present the optical and radio lightcurves, and argue
that in concert they provide clear evidence for evolution that can be
understood in the context of relatively simple jet models for the
ejecta.  The level of collimation implied for this event reduces, by a
factor $> 100$, the energy required to produce the gamma-ray flash.

\section{Observations}

GRB~990510, imaged by the {\em BeppoSAX} WFC on May~10.37 (UT)
(\cite{dad+99}), was a long ($\sim$75~s) relatively bright event with
a fluence (E $>$ 20~keV) of $2.6 \times 10^{-5}$~erg~cm$^{-2}$,
ranking it fourth among the SAX WFC localized sample, and in the top
10\% of BATSE bursts (\cite{kip+99,afc+99})\footnote{GCN circulars are
available at
http://lheawww.gsfc.nasa.gov/docs/gamcosray/legr/bacodine/gcn\_main.html.}.
After announcement of the WFC position by the SAX team, numerous
groups began the search for an optical transient (OT), eventually
discovered by Vreeswijk {\em et al.} (1999a)\nocite{vgr+99a}. The OT
is coincident with a fading X-ray source seen in the {\em BeppoSAX}
Narrow Field Instruments (NFI) (\cite{kha+99}). Spectra taken with the
VLT (\cite{vgr+99b}) identify numerous absorption lines, determining a
minimum redshift of $1.619 \pm 0.002$.  Adopting this as the source
redshift implies an isotropic gamma-ray energy release of $2.9 \times
10^{53}$~erg (we employ a standard Friedmann cosmology with $H_o =
65$~km~s$^{-1}$Mpc$^{-1}$, $\Omega_o = 0.2, \Lambda = 0$ throughout).

\begin{figure}
\clearpage
\centerline{\epsfig{file=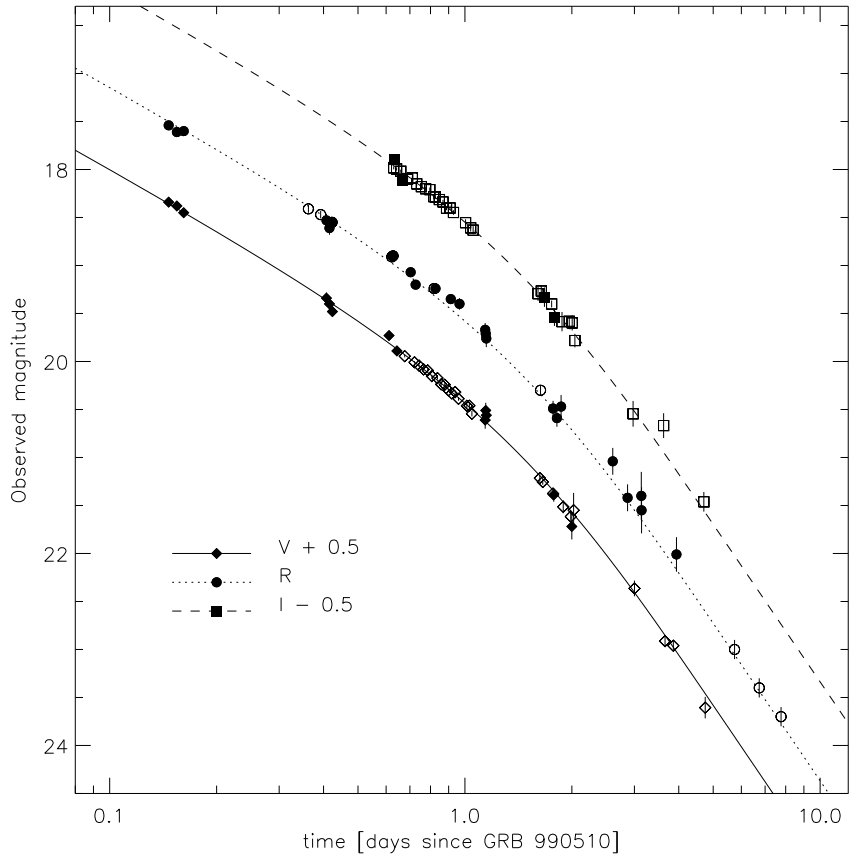,angle=90,width=12cm,angle=-90}}
\caption[optical.eps]{Optical light-curves of the transient afterglow of GRB
990510.  In addition to photometry from our group (filled symbols -- see Table
1), we have augmented the light curves with data from the
literature (open symbols).  The photometric zero-points in Landolt
$V$-band from our group are consistent with that of the OGLE group
(\cite{pu99b}) and the $I$-band zero-point is from the OGLE group.  Some
$R$-band measurements were based on an incorrect calibration 
of a secondary star in the field (\cite{gvr+99}) and we have recalibrated these
measurements.\label{fig-optical}}
\end{figure}

We commenced optical observations of the 3\arcmin\ {\em BeppoSAX} WFC
error circle using the Mount Stromlo 50-inch telescope 3.5~hr after
the event, and continued using in addition the Yale 1-m on Cerro
Tololo, and the 40-inch at Las Campanas.  Radio observations began at
the Australia Telescope Compact Array (ATCA), in Narrabri, Australia
about 17 hours following the GRB. Tables~\ref{tab-obsb}, \ref{tab-obsv},
\ref{tab-obsr} and
\ref{tab-obsi} present the BVRI optical data taken by our
collaboration (quoted errors are 1-$\sigma$ statistical
uncertainties). The VR and I lightcurves, along with points from
numerous other groups reported in the literature
(\cite{gvr+99,kgs+99,sgk+99,pu99a,pu99b,cfg+99,lcg99,pu99c,mil+99})
are plotted in Figure~\ref{fig-optical}. We have calibrated the
reported magnitudes to the Landolt bandpass system (approximately
Johnson-Cousins). For calibration, we observed a number of Landolt
Stars on May 11 under photometric conditions with the MSO 50-inch.
The uncertainty in the zero point of the calibration introduces a
magnitude error of $\pm 0.03$ in all bands.

\begin{table}[tb]
\begin{center}
\caption{B-band Photometry of 990510}
\label{tab-obsb}
\begin{tabular}{lcc}
\hline\hline
\multicolumn{1}{c}{Date in May} (UT)&   
\multicolumn{1}{c}{Magnitude$^a$} & Telescope \\ \hline
10.971 &  $19.86 \pm 0.05$ & Yale 1-m \\
11.058 &  $17.88 \pm 0.05$ & Yale 1-m \\
11.131 &  $17.95 \pm 0.05$ & Yale 1-m \\
11.154 &  $18.84 \pm 0.06$ & Yale 1-m \\
11.180 &  $18.90 \pm 0.06$ & Yale 1-m \\
11.207 &  $18.98 \pm 0.06$ & Yale 1-m \\
11.266 &  $19.23 \pm 0.06$ & Yale 1-m \\
11.292 &  $19.39 \pm 0.06$ & Yale 1-m  \\
11.320 &  $20.11 \pm 0.06$ & Yale 1-m \\
12.125 &  $20.01 \pm 0.08$ & Yale 1-m\\
12.171 &  $20.06 \pm 0.09$ & Yale 1-m \\
12.221 &  $20.89 \pm 0.09$ & Yale 1-m \\
12.300 &  $21.22 \pm 0.12$ & Yale 1-m \\
12.996 &  $21.22 \pm 0.17$ & Yale 1-m \\
\hline
\end{tabular}
\end{center}
\end{table}

\begin{table}[tb]
\begin{center}
\caption{V-band Photometry of 990510}
\label{tab-obsv}
\begin{tabular}{lcc}
\hline\hline
\multicolumn{1}{c}{Date in May} (UT)&   
\multicolumn{1}{c}{Magnitude$^a$} & Telescope \\ \hline
10.514 &  $17.84 \pm 0.02$ & MSO-50 \\
10.522 &  $17.88 \pm 0.02$ & MSO-50 \\
10.529 &  $17.95 \pm 0.01$ & MSO-50 \\
10.775 &  $18.84 \pm 0.06$ & MSO-50 \\
10.783 &  $18.90 \pm 0.08$ & MSO-50 \\
10.791 &  $18.98 \pm 0.05$ & MSO-50 \\
10.979 &  $19.23 \pm 0.04$ & Yale 1-m \\
11.011 &  $19.39 \pm 0.05$ & LCO-40  \\
11.508 &  $20.11 \pm 0.09$ & MSO-50 \\
11.512 &  $20.01 \pm 0.08$ & MSO-50 \\
11.516 &  $20.06 \pm 0.07$ & MSO-50 \\
12.146 &  $20.89 \pm 0.07$ & Yale 1-m \\
12.367 &  $21.22 \pm 0.14$ & LCO-40 \\
\hline
\end{tabular}
\end{center}
\end{table}

\begin{table}[tb]
\begin{center}
\caption{R band Photometry of 990510}
\label{tab-obsr}
\begin{tabular}{lcc}
\hline\hline
\multicolumn{1}{c}{Date in May} (UT)&   
\multicolumn{1}{c}{Magnitude$^a$} & Telescope \\ \hline
10.514 &  $17.54 \pm 0.02$ & MSO-50 \\
10.522 &  $17.61 \pm 0.02$ & MSO-50 \\
10.529 &  $17.60 \pm 0.02$ & MSO-50 \\
10.775 &  $18.53 \pm 0.07$ & MSO-50 \\
10.783 &  $18.61 \pm 0.07$ & MSO-50 \\
10.791 &  $18.55 \pm 0.04$ & MSO-50 \\
10.992 &  $18.90 \pm 0.04$ & Yale 1-m \\
11.071 &  $19.07 \pm 0.04$ & Yale 1-m \\
11.094 &  $19.20 \pm 0.04$ & Yale 1-m \\
11.194 &  $19.24 \pm 0.04$ & Yale 1-m \\
11.280 &  $19.35 \pm 0.05$ & Yale 1-m \\
11.333 &  $19.40 \pm 0.06$ & Yale 1-m \\
11.508 &  $19.67 \pm 0.07$ & MSO-50 \\
11.512 &  $19.71 \pm 0.06$ & MSO-50 \\
11.516 &  $19.76 \pm 0.09$ & MSO-50 \\
12.138 &  $20.49 \pm 0.08$ & Yale 1-m \\
12.183 &  $20.59 \pm 0.09$ & Yale 1-m \\
12.233 &  $20.47 \pm 0.12$ & Yale 1-m \\
12.975 &  $21.04 \pm 0.14$ & Yale 1-m \\
13.238 &  $21.42 \pm 0.14$ & Yale 1-m \\
14.308 &  $22.01 \pm 0.18$ & Yale 1-m \\ \hline
\end{tabular}
\end{center}
\end{table}

\begin{table}[tb]
\begin{center}
\caption{I band Photometry of 990510}
\label{tab-obsi}
\begin{tabular}{lcc}
\hline\hline
\multicolumn{1}{c}{Date in May} (UT)&   
\multicolumn{1}{c}{Magnitude$^a$} & Telescope \\ \hline
10.999 &  $18.40 \pm 0.04$ & Yale 1-m \\
12.154 &  $20.04 \pm 0.09$ & Yale 1-m \\
11.034 &  $18.61 \pm 0.05$ &LCO-40  \\    
12.042 &  $19.83 \pm 0.10$ &LCO-40  \\
\hline
\end{tabular}
\end{center}
\end{table}

From Figure~\ref{fig-optical}, it is evident that the lightcurve
steepens contemporaneously in all bands between day 1 and 2.
To characterize the shape, we fit the data with the following analytic
four-parameter function:
\begin{equation}
F_{\nu}(t) = f_* (t/t_*)^{\alpha_1} [1 - \exp{(-J)}]/J ;~~
J(t,t_*,\alpha_1,\alpha_2) = (t/t_*)^{(\alpha_1 - \alpha_2)}
\end{equation}
The functional form has no physical significance, but provides a good
description of the data, and has the property that the asymptotic
power law indices are $\alpha_1$ and $\alpha_2$ at early and late
times respectively.  Fitting the V,R, and I data (excluding B due to
larger statistical uncertainties) simultaneously
yields $t_* = 1.20 \pm 0.08$~days, $\alpha_1 = -0.82 \pm 0.02$, and
$\alpha_2 = -2.18 \pm 0.05$, where the errors are formal 1--$\sigma$
errors, and do not reflect the covariance between parameters. The
$\chi^2$ for the fit is acceptable: 65 for 82 d.o.f..  We have,
removed 5 out of the 92 total data points with uncertain calibrations.
Due to calibration uncertainty, we cannot determine if the lightcurve
exhibits variability on timescales shorter than the trend described by
the functional fit.  The difference in fit parameters from those found
by Stanek {\em et al.}  (1999)\nocite{sgk+99a} is due to the slightly
different function used.  Using the same function, we find consistency
with his results to better than 2-$\sigma$ in all parameters.

To derive the extinction-corrected normalizations, obtained by fitting
with the shape described above, we use the astrometric position from
Hjorth {\em et al.} (1999)\nocite{hbp+99} (RA = 13:38:07.11, Dec =
$-80:29:48.2$ (J2000)) and the dust maps from Schegel {\em et al.}
(1998)\nocite{sfd98}. The resulting Galactic extinction in the
direction of the transient is E(B-V) = 0.20.  In the standard Landolt
bandpass system, assuming $R_V = A_V/E(B-V) = 3.1$, we obtain $A_B$ =
0.87, $A_V$ = 0.67, $A_R$ = 0.54, $A_I$ = 0.40.
After correction, the magnitudes corresponding to the flux, $f_*$ in
Equation~1 are: $V_* = 19.03 \pm 0.01$, $I_* = 18.42 \pm 0.01$, $R_* =
18.81 \pm 0.01$. The errors are the formal 1--$\sigma$ errors from the
fit, with an additional $\pm 0.03$~mag due to the uncertain zero-point
calibration.


Observations of the field around GRB 990510 with ATCA began on 1999
May 10 at 22:36 UT. All observations (Table~2) used a
bandwidth of 128 MHz and two orthogonal linear polarizations for each
wavelength pair.  A radio afterglow is clearly detected, starting
$\sim$3 days after the event (Figure~\ref{fig-radio}).  The error bars
provided in the table are statistical (radiometric) errors only.  At
early times, variation due to interstellar scintillation will dominate
the error in flux determination from the source (see caption
Figure~2).

\section{Evidence for a Jet}

The majority of other well-studied GRBs, in particular GRB~970228 and
GRB~970508, have afterglow lightcurves that decay monotonically for
the first month or more, and these have been interpreted in the
context of spherical fireball models (e.g. \cite{tav97,wrm97,rei97,gps99}).
In the optical, spherical models with typical parameters predict flux
rising quickly (within hours) to a maximum value, $f_m$ (at time
$t_m$), after which it decays as a power law, $t^{-\alpha}$ with
$\alpha \sim 1$.  At later times, the decay becomes somewhat faster (a
change in $\alpha$ of 0.25), as the cooling break sweeps across the
band (\cite{spn98}).  In the radio band, above the self-absorption
frequency, the behavior is similar, but with typical values of $t_m
\sim 1$week.

The observed optical and radio decay of GRB~990510 is quite distinct,
showing frequency-independent steepening in the optical and early
decline in the radio on a timescale of 1 day; behavior clearly
inconsistent with spherical models. An achromatic break or steepening
in light curves is expected if the emitting surface has a
non-spherical geometry. At any given time, due to relativistic
beaming, only a small portion of the emitting surface with opening
angle 1/$\gamma$ is visible.  At early times, (when $\theta_o \gsim
1/\gamma$), the observed lightcurve from a collimated source is
identical to that of a sphere. As the fireball evolves and $\gamma$
decreases, the beaming angle will eventually exceed the opening angle
of the jet, and we expect to see a deficit in the emission -- i.e. a
break in the lightcurve. At a comparable or later time
(\cite{rho99,sph99,pm98}) the jet will begin to spread laterally,
causing a further steepening.

To model the lightcurve, we adopt the afterglow analysis for a jet
source given in Sari {\em et al.} (1999).  At early times ($\gamma >
\theta_o^{-1}$) the lightcurve is given by the spherical solution;
$F(\nu_o) \propto t^{\alpha}$ with $\alpha = -3(p-1)/4$ if the
electrons are not cooling, and $\alpha = -3p/4 + 1/2$ if they are.
From the GRB~990510 early time optical slope, $\alpha_1 = -0.82$, and
we derive $p = 2.1$ assuming the electrons producing the optical
emission are in the slow cooling regime, and $p=1.76$ otherwise.  The
latter value would result in the electron energy being unbounded, and
we conclude that $p=2.1$.  At late times ($\gamma <
\theta_o^{-1}$), when the evolution is dominated by the spreading of
the jet, the model predicts $\alpha = -p$, independent of the cooling
regime. Indeed, our measured value of $\alpha_2 = -2.18 \pm 0.05$ is
consistent with this expectation.

The optical data allow us to infer $p$ and the epoch of the break
(related to the opening angle of the jet).  However, in order to fully
characterize the afterglow we also need to determine: (a) $\nu_a$, the
self absorption frequency, (b) $F_m$ and $t_m$ and (c) the cooling
frequency, $\nu_c$ at a given epoch.  The optical observations show
that even at early times the optical flux is decaying, and is
therefore above $\nu_m$. The radio, however, is well below $\nu_m$,
and by combining the ATCA and optical data we can derive $F_m$, $t_m$,
and $\nu_m$.  Following Sari et al (1999), we have fitted a $t^{-1/3}$
powerlaw to the four radio points and obtained $F_{8.7{\rm GHz}} \cong
204 \mu{\rm Jy} (t/t_1)^{-1/3}$, where $t_1=3.3$d is the time of the
second radio detection. Using this and the optical data at $t_1$ we
get $\nu_m(t_1)=280$~GHz and $F_{m}(t_1)=650~\mu$Jy.  After the jet
begins to spread, $\nu_m$ decays as $t^{-2}$, and we expect $\nu_m$ to
arrive at radio frequencies at $\sim 19$ days, producing a break in
the radio lightcurve to the $t^{-p}$ slope seen in the optical.  In
the above, we have assumed that $\nu_a$ is below 8.7~GHz. A $\chi^2$
analysis constrains the 4.8 -- 8.7~GHz spectral slope to be between
$-1.3$ and $0.4$ (95\% confidance), consistent with the $\nu^{1/3}$
slope expected if $\nu_a < 8.7$~GHz, and inconsistent with the
$\nu^{2}$ expected if $\nu_a > 8.7$~GHz. 

\begin{table}
\begin{center}
\begin{tabular}{|l|c|c|c|c|}
\hline
Date in   &    Freq.  & Flux density  &  Integration        &   Angular Res. \\
May (UT)    &    (GHz)  &    (uJy)      &      (hrs)        &   (arcsec)     \\
\hline\hline
11.09   &   4.8     &  $110 \pm 69$  &  7.5         &     4.2 $\times$ 1.8  \\
11.09   &   8.6     &  $104 \pm 74$  &  7.5         &     1.9 $\times$ 1.3  \\
13.68   &   8.7     &   $227 \pm 30$  &  9.0         &     1.9 $\times$ 1.3  \\
15.61   &   8.7      &   $202 \pm 31$  &  8.0         &     1.8 $\times$ 1.4  \\
17.58   &   8.7     &  $138 \pm 32$  & 6.6      &  2.1 $\times$ 1.2 \\    
19.59   &  4.8      &   $177 \pm 36$ &  11.4      & 3.1 $\times$ 2.6 \\
19.59    &  8.6      & $127 \pm 31$   &  11.4      & 1.7 $\times$ 1.5  \\
25.32    & 8.7       & $82 \pm 32$   &   10.6   &    2.2 $\times$ 1.2   \\
46.81   & 8.7        & $-1 \pm 28$   &  11.7   & 4.0 $\times$ 3.6   \\ 
\hline
\end{tabular}
\caption{ATCA Radio flux measurements. The date indicates the observation center. \label{tab-radio}}
\end{center}
\end{table}

\begin{figure}
\centerline{\epsfig{file=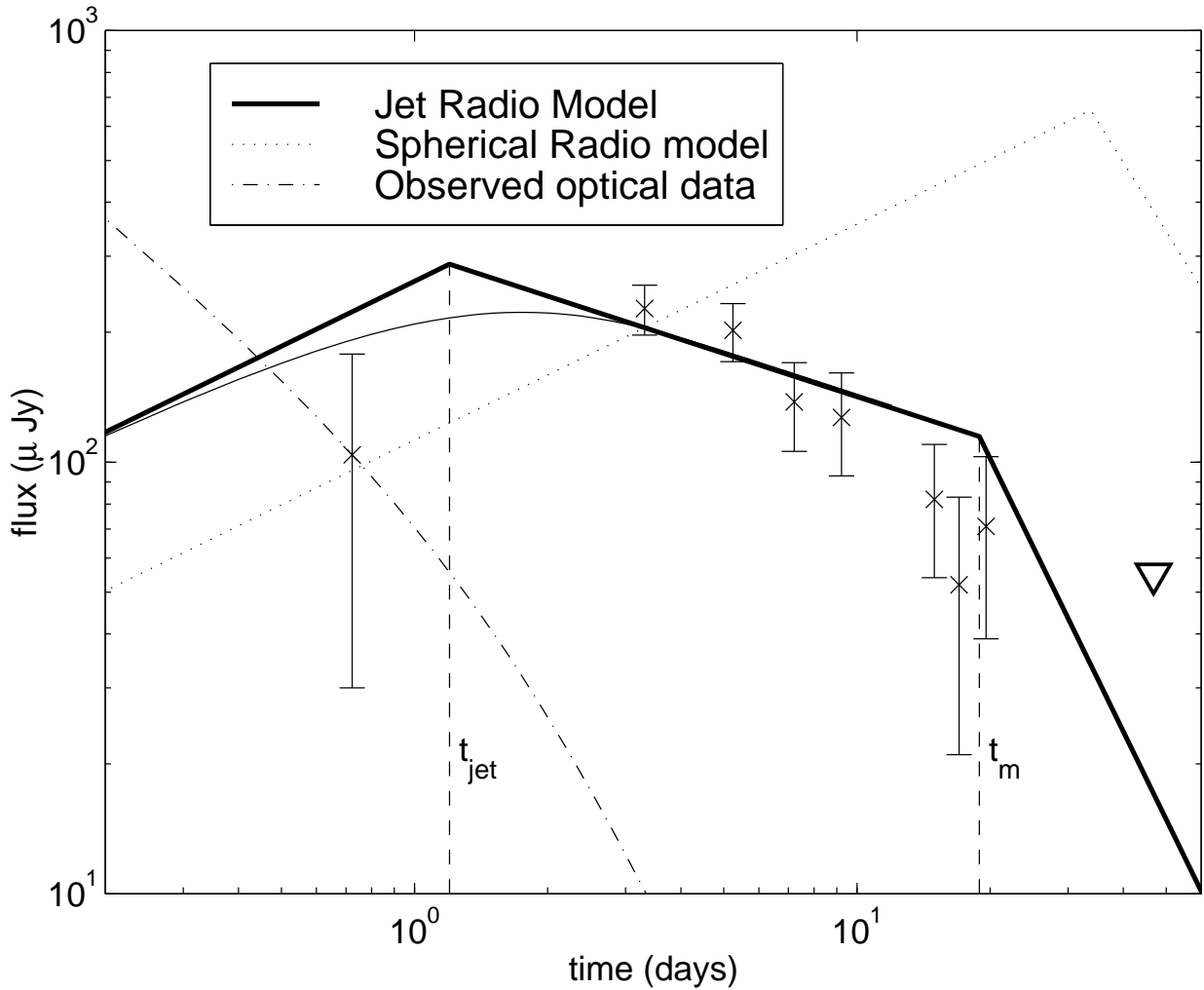,angle=0,width=\columnwidth,angle=0}}
\figcaption[radio.eps]{Observed and predicted radio lightcurves at 8.6 GHz. 
Detections are indicated by the crosses, with error bars indicating
the rms noise in the image. The true flux uncertainty is dominated by
the signal modulation due to refractive interstellar scintillation
(e.g. Frail et al. 1997). Using the Galactic scattering model of
Taylor \& Cordes (1993), and the formalism from Goodman (1997), we
calculate a scintillation timescale of 2 hrs in the first few weeks
after the burst. Although our typical 8 hour integrations
average over the scintillation, we expect modulation of the mean
flux density of order 50\%. Predictions for the evolution of the radio
flux density (solid line) are based on the jet model of Sari et
al. (1999) (see text for more details). The dotted line shows the
model prediction for a spherical fireball. The dotted-dashed line
illustrates the observed optical behavior.\label{fig-radio}}
\end{figure}

Figure~\ref{fig-radio} shows the radio lightcurve along with the
prediction for both spherical (dotted line) and collimated (solid)
ejecta.  The relatively sharp transition in the GRB~990510 decay to the
asymptotic value $\alpha_2 = -p$ expected when both the jet edge becomes
visible and when lateral spreading begins suggest both transitions occur
at similar times in this event.

Using the gamma-ray energy of $2.9 \times 10^{53}$~erg, we find a
Lorentz factor at the jet break time of $12
(n/1$cm$^{-3})^{-1/8}$. This implies an opening angle of $\theta_o
= 0.08 (n/1$cm$^{-3})^{1/8}$, and for a two-sided jet the energy is
reduced by a factor $2/\theta_o^2 \cong 300$, to $1
\times 10^{51}(n/1$cm$^{-3})^{1/4}$~erg
\footnote{The estimates of Rhoads (1999) will give a smaller opening angle and
therefore a lower energy, here we have used the estimates in Sari et
al.  (1999).}.

\section{Conclusion}

With one of the best-sampled optical lightcurves, and simultaneous
early time radio observations, GRB~990510 provides the clearest
signature observed to-date for collimation of the ejecta in GRB
sources.  The achromatic steepening in the optical lightcurve, as well
as the early decay, after $t \sim 1$~day, of the radio emission is
inconsistent with other observed afterglows that have been modeled
with spherically-symmetric ejecta.  The GRB~990510 afterglow emission
can be remarkably well fit by a simple model for the jet evolution.

It is interesting to ask if the observations to-date are consistent
with all GRB engines having an energy release of $\lsim
10^{52}$~erg, with the wide observed luminosity distribution being due
to variation in the degree of collimation.  Of GRBs with measured
redshifts for which the gamma-ray energy release can be calculated,
only GRB~990123 and GRB~990510 show breaks in the optical lightcurves
on timescales less than 1 week, and interestingly these are among the
highest fluence SAX events to-date. GRB~990123 has an implied
isotropic energy release of $3.4
\times 10^{54}$~erg, which reduces by a factor $\sim 100$ if the
lightcurve break occuring at $t \sim 2$ days is interpreted as the
signature of a jet.  As argued here, the energy required for
GRB~990510 in the context of the jet model is $\sim 10^{51}$~erg.  In
contrast, 970508 and 970228 show no evidence for a jet in the optical
(although 970508 may in radio), however their isotropic energy release
is quite modest: only $8 \times 10^{51}$~erg and $5
\times 10^{51}$ respectively.    The candidates for the largest energy release; highest
gamma-ray fluence where no evidence for collimation is seen are
GRB~971214 ($z=3.2$) with $E_{\gamma} = 3 \times 10^{53}$~erg
(\cite{kdr+98}) and GRB~980703 ($z=0.966$) $E_{\gamma} = 1 \times
10^{53}$~erg (\cite{d+99}).  Lightcurve observations of these events
are, however, limited to $t \lsim 2$~weeks, and so collimation may
still reduce the energy of these bursts by factors of $\sim$40, still
consistent with a total energy release $\lsim 10^{52}$~erg.

\nocite{tc93}
\nocite{goo97}

\acknowledgements{We thank Scott Barthelmy for operating
the GCN, and the staffs of LCO, MSSSO, and ATCA, and the entire
BeppoSAX team. This work was supported by grants from NSF (SRK,SGD),
NASA (FAH, SRK), and the Bressler Foundation (SGD). 


\end{document}